\documentclass{iopart}
\usepackage{epsfig}
\usepackage{graphicx}
\usepackage{cite}

\newcommand{\nc}{\newcommand}
\nc{\be}{\begin{equation}}
\nc{\ee}{\end{equation}}
\nc{\bea}{\begin{eqnarray}}
\nc{\eea}{\end{eqnarray}}
\nc{\disp}{\displaystyle}
\nc{\ade}{\mbox{$A$-$D$-$E$}}
\nc{\calN}{{\cal N}}
\nc{\calC}{{\cal C}}
\nc{\calM}{{\cal M}}
\nc{\calS}{{\cal S}}
\nc{\phit}{\hat{\varphi}}
\nc{\chit}{\hat{\chi}}
\nc{\hcalN}{\hat{\calN}}
\nc{\hcalS}{\hat{\calS}}
\nc{\hS}{\hat{S}}
\nc{\sigmad}{\sigma^\dagger}
\nc{\psid}{\psi^\dagger}

\def\sstyle{\scriptstyle}

\begin{document}
    \title[Conformal invariance and its breaking]{Conformal invariance and its breaking in a stochastic model of a
fluctuating interface}

\author{Francisco C. Alcaraz$^1$, Erel Levine$^2$ and
Vladimir Rittenberg$^{3,4}$}\address{$^1$Instituto de F\'{\i}sica de
S\~ao Carlos, Universidade de S\~ao Paulo,Caixa Postal 369, 13560-590 S\~ao Carlos, SP, Brazil\\
$^2$Center for Theoretical Biological Physics, University of
California at San Diego, La Jolla, CA 92093\\
$^3$Department of Mathematics and Statistics, University of Melbourne, Parkville, Victoria 3010, Australia \\
$^4$Physikalisches Institut, Bonn University, 53115 Bonn, Germany.}

\date{\today}

\begin{abstract}
 Using Monte-Carlo simulations on large lattices, we study the effects of
 changing the parameter $u$ (the ratio of the adsorption and desorption
 rates) of the raise and peel model. This is a  nonlocal
 stochastic model of a fluctuating interface. We show that for $0<u<1$ the
 system is massive, for $u=1$ it is massless and conformal invariant. For $u>1$
 the conformal invariance is broken. The system is in a scale
 invariant   phase with the dynamical critical exponent $z$ and other 
critical exponents varying continuously with $u$.
 As far as we know it is the
 first example of a system which shows such a  behavior. Moreover in the
 broken phase, the critical exponents vary continuously with the parameter
 $u$. This stays true also for the critical exponent $\tau$ which characterizes
 the probability distribution function of avalanches (the critical exponent $D$
 staying unchanged).
\end{abstract}
%\pacs{02.50.Ey,\ 11.25.Hf,\ 05.50.+q,\ 75.10.Hk}

\section{Introduction}

 The raise and peel model (RPM) of a fluctuating one-dimensional
interface was introduced in Ref.~\cite{GNPR}. In this model, the
evolution of the interface (described in terms of RSOS paths), in the
presence of a rarefied gas of tiles (tilted squares or particles),
follows Markovian dynamics. When a tile hits a site of the interface
it can be reflected, adsorbed with a probability $u_a$ ("raising"
the interface) or trigger a desorption process with a probability
$u_d$, depending on the heights of the two neighbouring sites. In
the desorption process part of the top layer of the interface
evaporates (the interface is "peeled"). The number of tiles which
leave the interface in one desorption process, which may be
macroscopic, define the size of an avalanche.

The model depends on a single parameter $u = u_a/u_d$ or alternatively $w = 1/u$. The case
$u = 1$ was discussed  in detail in \cite{GNPR} in the continuum time limit.
 There are two reasons why this case has received so much attention:

a) The time evolution of the system is given by a Hamiltonian $H$ which, in
the finite-size scaling limit, has a known spectrum. The reason is that
in this limit $H$ describes a $c=0$ conformal field theory (CFT) ($c$ is
the central charge of the Virasoro algebra)  and the spectrum is given by
characters of this algebra \cite{GNPR1}. Moreover this observation implies
that the dynamical critical exponent is $z = 1$.

b) The weights of the various RSOS paths in the probability
distribution function (PDF) of the stationary state for a finite system are
related to a counting problem of alternating sign matrices with different
topologies (see \cite{GNPR2} and references therein). Alternating
sign matrices have miraculous combinatorial properties\cite{BRES} and this has
allowed to guess (conjecture) the expressions of several
physical quantities in the stationary state for any number of
lattice sites based on exact calculations for small lattices.

 In \cite{GNPR} very little work was done for $u$ different of one except for a crude
estimate of the phase diagram. The reason is simple: for $u=1$ one
has used the exact knowledge of the stationary state for lattice
sizes up to 18 sites (it is hard, although possible, to go beyond
this number) and made  (unproven) conjectures for any
lattice sizes. For $u\neq1$, one  has only diagonalized numerically
 (up to 14 sites) the Hamiltonian, that gives the time evolution of the 
system,  in order to estimate in which
domain of $u$ the system is massless (scale invariant) or massive
(finite correlation lenghts). The results were inconclusive.

 In this paper we present an extensive study of the model using Monte
Carlo simulations. What was our motivation?

 a) For $u=w=1$, the RPM is, to our knowledge, the first example of a
stochastic process in which conformal invariance, a time-space symmetry, 
 plays a role and one is
interested to see some consequences. 
 Let us stress that this is 
different of the one time- two space abelian sand-pile model where conformal 
invariance is seen in the stationary state \cite{DHA}
We also wanted to check the conjectures made
in \cite{GNPR} for large lattice sizes and to study other properties of the model.

b) When $u=1/w  \neq 1$ several scenarios are possible. The system may stay
conformal invariant, may get massive or stay scale invariant without
being conformal invariant. According to the standard lore (see Refs.~\cite{POL}
and
\cite{RC} for a more precise definition of the conditions) provided that the
interactions are local, scale invariance implies conformal invariance.
Although for $u=w=1$, the RPM can be mapped onto the spin zero sector of an
$U_q(sl(2))$ invariant XXZ quantum chain \cite{GNPR} in which the interaction is
local, as far as we can see, for $u \neq 1$ a mapping onto a system with local
interactions is not possible. Therefore, keeping in mind that the
desorption process is not local, for $u \neq 1$  
we still have the interesting possibility that the system can be scale invariant
without being conformal invariant.

 We will show that for  $u<1$ the system is massive while for $u>1$ ($w<1$) the
 system is scale invariant but not conformal invariant, with the critical 
exponent $z$ \cite{hohenberg} and other  critical 
exponents varying 
  continuously with $w$. At $w=0$ the system becomes massive again. We think that it is for the first time,
 that such a phenomenon is seen.

 c) As noticed already in Ref.~\cite{GNPR} and confirmed in the present paper,
 for $u=1$ the probability distribution function (PDF) of avalanches 
  triggered by  tiles hitting the
 interface in the stationary state, has a long tail, the critical exponents
 being $D=1$ and $\tau=3$ (see Section~5 and Ref.~\cite{TMS} for definition of the
 exponents),
 which means that the model exhibits self-organized criticality (SOC). We will
 show how the PDF of avalanches behave if one varies $w$.

  The paper is organized as follows.
  In Section~2, following \cite{GNPR} we
  present the model and define the quantities which we found useful to
  characterize the interface. Those are contact points, clusters, heights
  and the  fraction of sites of the interface where desorption does not take
  place (FND), {\it i.e.} the density of local maxima or minima.

  In Section~3 we describe the properties of the stationary states for
   various values of $u$, and derive the phase diagram of the model 
  (Fig.~\ref{fig:phase1}). We give the variation of the number of clusters with
   the size of the system $L$. We show that for $u<1$ the density of clusters is
   finite for large values of $L$. For $u=1$ this density vanishes 
  like $L^{-\frac{1}{3}}$. 
    For $w<1$
   the density of clusters also vanishes as  a power law with an exponent
   which varies with $w$. We also give the local density of contact points as a
   function of the size of the system and its finite-size scaling behavior.
   We  discuss the size dependence of the average height.
   The variation of the FND with $w$ and the system sizes is given in
   Section~5.

  In Section~4 we give the values of the dynamical critical exponent $z$ for
    various values of $w$. We estimate $z$ using the Family-Vicsek scaling for
    various observables. We show that $z$ decreases from the value $1$ for $w=1$
    towards $0$ at small values of $w$. For $w=1$ we show how the knowledge of the
    finite-size  spectrum (explained in appendix A)
    gives  information on the
    Family-Vicsek scaling functions. We also give the two-contact points
    correlation function obtained using Monte-Carlo simulations and
    illustrate the predictions  of conformal invariance.

   In Section~5 we study the PDF of avalanches for different  values of $w$.
   The average size of the avalanches stays finite in the whole domain
   $1 > w>0$ and its value is given remarkably well by mean field.
   We
     show that the exponent $D$ is equal to 1 for all values of $w$
     but that the
     exponent $\tau$ \cite{TMS} changes from the value 3 at $w=1$ to a  value close to
       2 for smaller values of $w$.
      Our conclusions are presented in Section~6.

\section{The raise and peel model}
\label{se:modeldef}

We consider a one-dimensional lattice with $L+1$  
($L=2n$) sites.  An interface is   formed by attaching at each site 
non-negative integer heights $h_i$ ($i=0,1,\ldots,L$) which obey the restricted
solid-on-solid (RSOS) rules:
\be
h_{i+1} - h_i = \pm1,\qquad h_0 = h_L = 0,\qquad h_i \geq 0.
\label{eq:heights}
\ee

\begin{figure}[t]
\centerline{
\begin{picture}(160,40)
\put(0,0){\epsfxsize=160pt\epsfbox{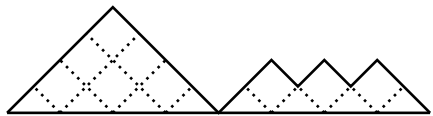}}
\end{picture}}
\caption{A configuration of the interface with three contact points and
two clusters for the lattice size $L=16$.}
\label{fig1}
\end{figure}

There are $C_n= (2n)!/((n+1)(n!)^2)$ possible configurations of the
interface. In Fig.~\ref{fig1} we show a configuration for
$n=8$ ($L=16$).
Alternatively, one can describe the interface using slope variables
$s_i = (h_{i+1} - h_{i-1})/2$,  ($i=1,...,L-1$).

The dynamics of the interface is described in a transparent way in the
language of tiles (tilted squares) which cover the area between the
interface and the substrate $(h_{2i}=0$, $h_{2i+1} =1$,
$(i=0,...,n))$ (see Fig.~\ref{fig1}).

We consider the interface separating a film of tiles
deposited on the substrate from a rarefied gas of tiles. We are interested
to find the evolution of the interface towards the stationary state and to
study the properties  of the interface in this state.

\begin{figure}[t]
\centerline{
\begin{picture}(160,80)
\put(0,10){\epsfxsize=160pt\epsfbox{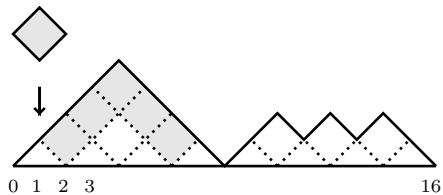}}
\put(-2,0){$\sstyle 0$}
\put(7,0){$\sstyle 1$}
\put(17,0){$\sstyle 2$}
\put(27,0){$\sstyle 3$}
\put(154,0){$\sstyle 16$}
\end{picture}}
\caption{A desorption event. The incoming tile at site $1$ triggers, 
with probability $u_d$,  an
avalanche of $5$ tiles, which are shaded. All of the shaded tiles are
removed in the desorption event.}
\label{fig2}
\end{figure}

The evolution of the system in discrete time (Monte-Carlo steps) is
given by the following rules. With a probability $P_i= 1/(L-1)$ a tile from
the gas hits site $i,\; (i=1,...L-1)$. Depending on the value of the slope
$s_i$ at the site $i$, the following processes can occur:
\begin{itemize}
\item[i)] $s_i=0$ and $h_i > h_{i-1}$. %(local maximum)

The tile hits a local peak and is reflected.
\item[ii)] $s_i=0$ and $h_i < h_{i-1}$. %(local minimum)

The tile hits a local minimum. With a probability $u_{\rm a}$ the tile
is adsorbed ($h_i \mapsto h_i+2$) and with a probability $1-u_{\rm
a}$ the tile is reflected.
\item[iii)] $s_i=1$.

With probability $u_{\rm d}$ the tile is reflected after triggering
the desorption of a layer of tiles from the segment
($h_j>h_i=h_{i+b},\; j=i+1,\ldots, i+b-1$),
{\it i.e.} $h_j \mapsto h_j-2$ for $j=i+1,...,i+b-1$. This layer contains
$b-1$ tiles (this is always an odd number). With a probability $1-u_{\rm d}$, the
tile is
reflected and no desorption takes place. For an example see Fig.~\ref{fig2}.
\item[iv)] $s_i=-1$.

With probability $u_{\rm d}$ the tile is reflected after triggering
the desorption of a layer of tiles belonging to the segment
($h_j>h_i=h_{i-b},\; j=i-b+1,\ldots, i-1$),
{\it i.e.}  $h_j \mapsto h_j-2$ for $j=i-b+1,...,i-1$. With
a probability $1-u_{\rm d}$ the tile is reflected and no desorption
takes place.
\end{itemize}

     In the Monte Carlo simulations we
     choose $u_d=1$ and $u_a=u$ for $u=u_a/u_d \leq 1$, and $u_d =w, u_a=1$  for $u>1$.
 For convenience we will use either $u$ or $w$ as the free 
parameter of the problem. We use these notations since we consider
below either very small values of $w$ (or $u$) and our intuition can
be helped by the fact that for $u=0$ or $w=0$ the properties of the
system are known. Namely, for both $u=0$ and $w=0$ the spectrum is
massive \cite{GNPR}, the stationary states being the substrate ($u=0$) and a
full triangle ($w=0$).

 It is useful to define some quantities which characterize the surface.
An obvious geometric observable
 is the set of sites $j$ ($j$ even) for which $h_j=0$
(the sites $0$ and $L$ always belong to this set). These sites are
also called contact points. This set is important, for example,  to
study  desorption. Desorption events are limited to the area between
two contact points, defined as a cluster (see Fig.~\ref{fig1}). As
is going to be shown in Section~3, the density of contact points gives
a local observable for which one can define various correlation
functions (see also Section~\ref{sec4}).

 We first define quantities in the stationary state. The average number of
 clusters $k(L)$ and the density of clusters $\rho(L)$ are defined by:
\be
k(L) = \langle \sum_{j=1}^{L} \delta_{h_{j},0}\rangle ,
\label{eq:clusterdef}
\ee
\be
\rho(L) = k(L)/L.
\label{eq:rhodef}
\ee
  Notice that the maximum value of $\rho(L)$ is $\frac{1}{2}$
  which is obtained when $u=0$
  and one has only one configuration in the system - the substrate.
   The average height is defined as:
\be
h(L) = \frac{\langle \sum_{j=1}^{L} h_j\rangle}{L}.
\label{eq:defh}
\ee
The lowest value of $h(L)$, corresponding to the substrate, is $\frac{1}{2}$.
    It is useful to consider also the average height in the middle of the
    system:
\be
h_{\frac{1}{2}}(L) = \langle h_{\frac{L}{2}}\rangle .
\label{eq:defhl2}
\ee
     A relevant quantity is the average of the fraction of the interface where
     desorption does not take place (FND)
\be
n(L) = \frac{1}{L-1}\langle \sum_{j=1}^{L-1}(1-|s_j|)\rangle .
\label{m}
\ee
      This quantity will be used in Section~5  since it allows to estimate the
      average number of  tiles desorbed in avalanches.
       The large $L$ behavior of any of these quantities, say $h(L)$, will be
       denoted $h_{\infty}$:
\be
 h_{\infty}\equiv  \lim_{L\rightarrow \infty}  h(L).
 \label{eq:defhas}
 \ee

        In Section~4 we are going discuss average quantities which are
    not only $L$ dependent but also time dependent. For these
    quantities we use, for example,  the notation $k(t,L)$ to denote the 
number of clusters.

\section{ The stationary states of the RPM}
\label{sec3}

%F33333333333333333333333333333333%____________________________________$
%
\begin{figure}[t]
\centerline{
\begin{picture}(260,190)
\put(0,0){\epsfxsize=260pt\epsfbox{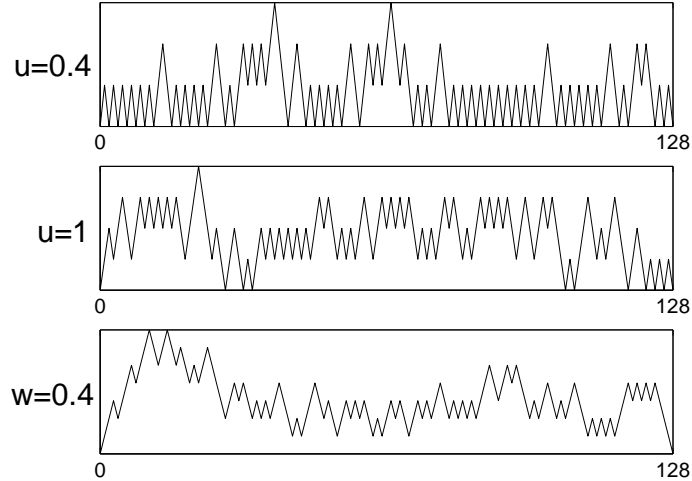}}
\end{picture}}
\caption{
  Typical configurations in the stationary states for
$ L= 128$ sites and three  values of~$u$.}
\label{fig3}
\end{figure}
%

%F3333333333333333333333333333333%________________________________________$

 In order to get an intuition on the behavior of the model when $u$
 changes, in Fig.~\ref{fig3} we show for $L =  128$ typical configurations in
 the
 stationary states for three values of the parameter $u$.
   One notices that for $u = 0.4$ there are many clusters,
  fewer for $u = 1$ and
  a single one for $w = 0.4$
(other, less probable, configurations do exhibit several clusters). We also notice that the average heights are quite small.
   The Monte Carlo simulations were done for various lattice sizes up to
   $L = 65536$ sites.

\begin{figure}[t]
\centerline{
\begin{picture}(300,50)
\put(0,15){\epsfxsize=300pt\epsfbox{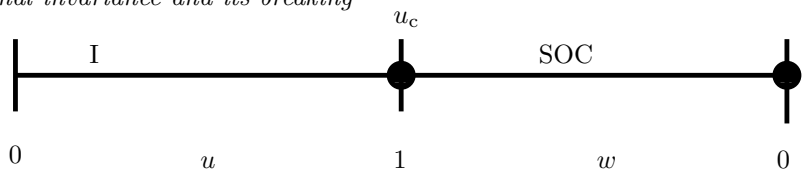}}
\put(-0.5,2){$0$}
\put(290,0){$0$}
\put(145,0){$1$}
\put(72,0){$u$}
\put(222,0){$w$}
%\put(270,40){$w_{\rm c}$}
\put(145,55){$u_{\rm c}$}
\put(30,40){I}
\put(200,40){SOC}
%\put(290,30){II}
\end{picture}}
\caption{
     Phase diagram of the raise and peel model. For $0 \leq u < 1$ (phase I),
     the
    model is massive. At $u = u_c=1$ it is massless and conformal invariant.
    For
    $1 > w >0 $ it is scale invariant with varying critical exponents and
    exhibits  SOC.
     For $w = 0$ the system is massive. }

\label{fig:phase1}
\end{figure}

    Before presenting our results, in Fig.~\ref{fig:phase1}
    we show what be believe is the
    phase diagram of the model.
     In the domain $0 < u< 1$ the system is massive undergoing a
    second order phase transition at $u = u_c = 1$. For $1 > w > 0$
    the conformal symmetry seen at
    $w= 1$ is broken but the system stays scale invariant. Since in this
    domain one has
    avalanches we call this phase a SOC phase. Another phase transition
    occurs at $w = 0$
    where the system becomes massive again.
    In the present section and in the next two
    ones we will give arguments which favor our conclusion.

%F4444444444444444444444444%    _________________________________________________________________________

%F44444444444444444444444444%    ______________________________________________________________________

\begin{figure}[t]
\centering{\includegraphics [angle=0,scale=0.39] {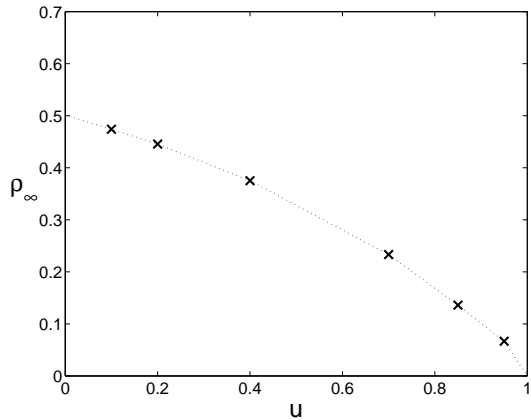}}
\caption{
        The density of clusters $\rho_{\infty}$ for various values
       of $u <  1$.}
\label{fig5}
 \end{figure}

     We first consider the density of clusters $\rho (L)$ in the domain
     $0\leq  u < 1$. Taking large values of $L$, one obtains the $u$
     dependence of
     $\rho_{\infty}$ shown in Fig.~\ref{fig5}. In all this domain the density
     of clusters
     stays finite. The density of clusters decreases from its maximum
     possible value $1/2$ which corresponds to the substrate at $u = 0$
     to the value
     zero for $u = 1$. This indicates that there is a phase transition
     at $u = 1$. From our Monte Carlo simulations it was hard to obtain the
     critical exponent which gives $\rho(u)$ when $u$ approaches the value
     one.

%F555555555555555555555555555

%F555555555555555555555555555555555

\begin{figure}[t]
\centering{ \includegraphics [angle=0,scale=0.46]
{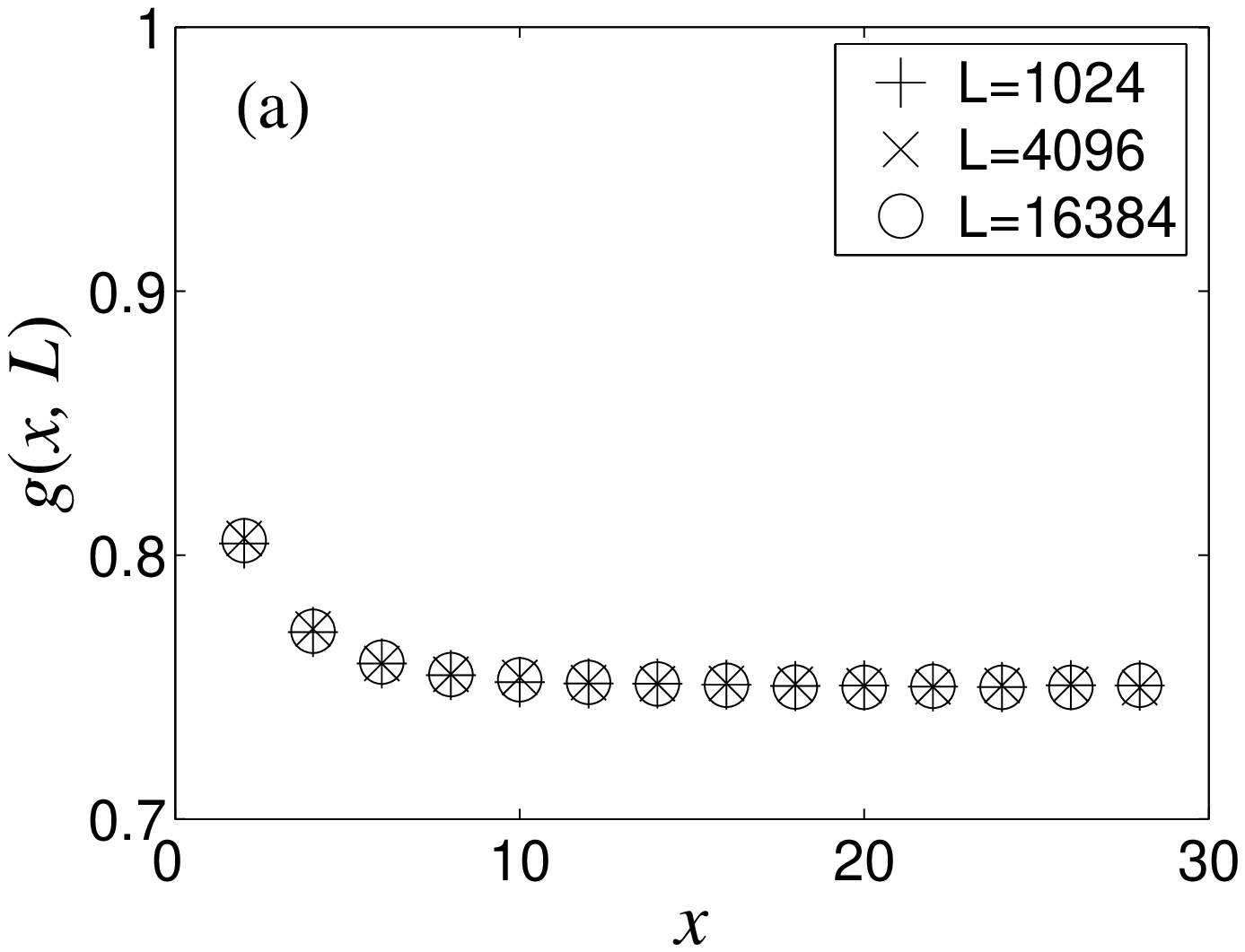}\hspace{-3.0cm}\hfill\includegraphics
[angle=0,scale=0.35] {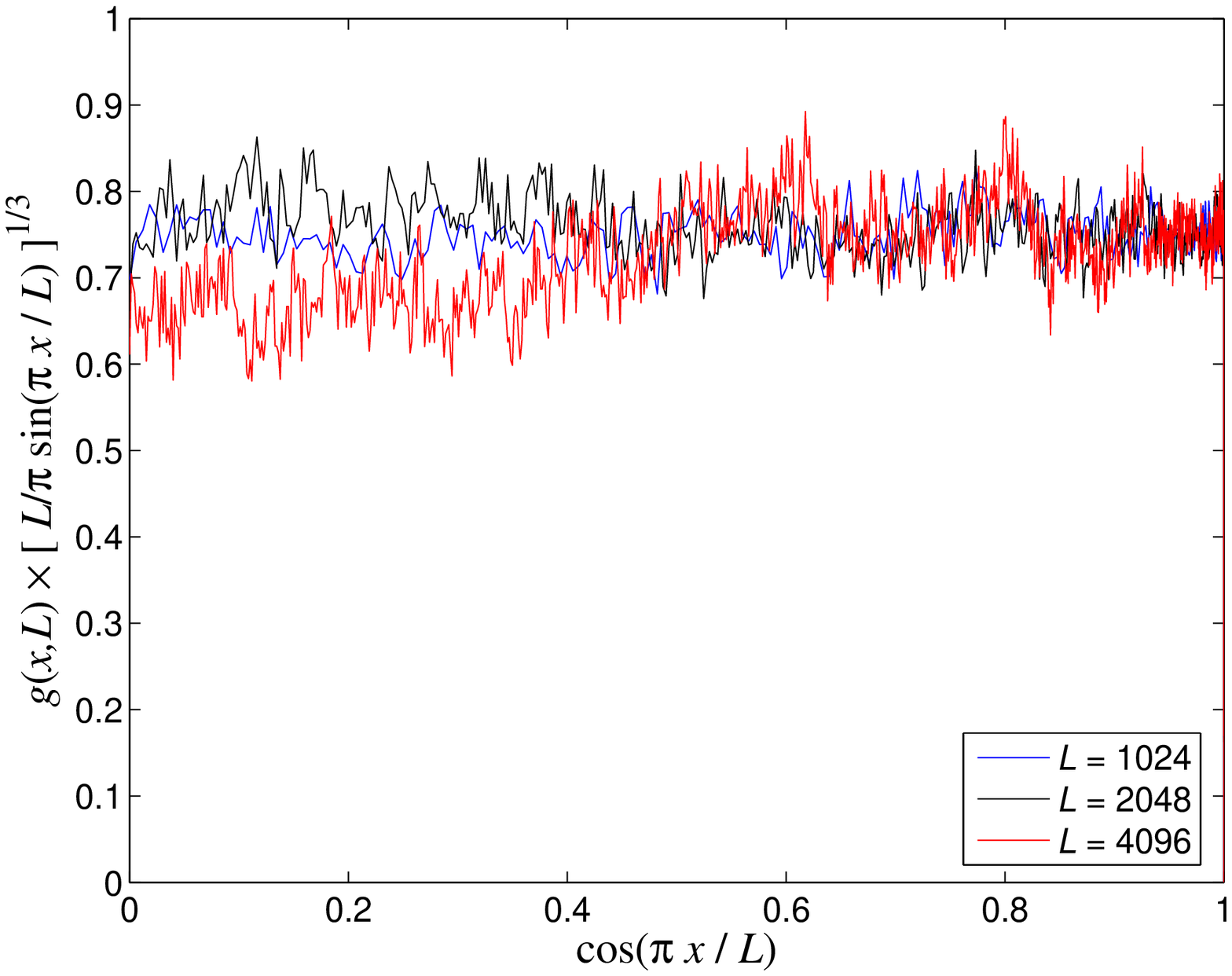} } \caption{
     (a) The local density of contact points  $g(x,L)$ at $u= 0.4$ for
    different lattice sizes $L$. (b) The scaling function 
$C(\cos (\pi x/L))$ 
defined in (\ref{n1}) for $u=1$.}
\label{fig6}\label{fig7}
 \end{figure}

        The local density of contact points $g(x,L)$ for $u = 0.4$ and different lattice sizes is shown
    in Fig.~\ref{fig6}(a). One
    observes that for small values of $x$, and large values of $L$,
    $g(x,L)$ decreases
    exponentially from the value 1 at $x = 0$ (not shown on Fig.~\ref{fig6}(a))  to a constant value
    in the bulk \footnote{This value is twice that seen in Fig.~\ref{fig5}, 
since for even $x$ one always have $g(x)=0$.}.

%   F666666666666666666666666

%_______________________________________________________________________________
%

%   F66666666666666666666666666

    If we keep in mind that for $u = 0$ one can diagonalize the
     related Hamiltonian of the stochastic process and
     show rigorously \cite{GNPR} that the system is massive, we can
     expect that
     the system is massive in the whole domain $0\leq u < 1$.

      We turn now to the interesting case $u = 1$. In Ref.~\cite{JANG} a
      conjecture was
      made which gives the probability to have $k$ clusters in a system
      of size $L$.
 For our purpose, we need only the behavior of the
      average number of clusters (see \cite{GNPR} Eq.(47)) for large system
      sizes, given by
\be 
\label{eq8} k(L) \simeq \frac{\Gamma(1/3) \sqrt{3}} {2\pi}
L^{\frac{2}{3}}, 
\ee
      which shows that the density of clusters vanishes algebraically for
      $u = 1$.
      We have checked this conjecture using Monte Carlo simulations up
      to a system
      of size $L=512$.    Since, as we show below, the density of clusters also vanishes
      algebraically  in the entire domain $w<1$, it is convenient to define a critical exponent
      $\alpha < 1 $
 which gives the power law increase of the number
      of clusters with the system size, $k(L) \sim L^\alpha$.
      For $u = 1$ we have obviously $\alpha = 2/3$.

We have considered, again for $u = 1$, the density of contact points $g(x,L)$
for several lattice sizes and noticed that for the function
\be \label{eq9}
G(x,L) = L^{\frac{1}{3}} g(x,L) = G(y),
\ee
where $y = x/L$ one has a data collapse. We make an {\it ansatz} about the
functional behaviour of $G(y)$ taking:
\be \label{n1}
 g(x,L) = [\frac{L}{\pi}\sin(\pi x/L)]^{-1/3} C(\cos (\pi x/L)).
\ee

The motivation of this {\it ansatz} will become clear in the next lines. 
In Fig.~\ref{fig6}(b) 
one shows the function $C(\cos(\pi x/L)$. It is practically a constant. If
one assume that $C$ is independent of $\pi x/L$, one can get its value by
integrating $g(x,L)$ (Eq. (\ref{n1})) and compare the result with the average
number of clusters given by (\ref{eq8}). One obtains:
\be \label{n2} 
C = -\frac{\sqrt 3}{6\pi^{\frac{5}{6}}}\Gamma(-\frac{1}{6}) = 
0.753149...,
\ee
in very good agreement with the data (see Fig.~\ref{fig6}(b)).

 If the density of contact points corresponds to a local operator in
conformal field theory, one expects \cite{BUXU},
\be \label{n3}
 g(x, L) = C [\frac{L}{\pi}\sin(\pi x/L)]^{X},
\ee
where $X = \Delta + \bar{\Delta}$  is the scaling dimension of the local
operator ($\Delta-\bar{\Delta}$  is the conformal spin). Comparing 
Eqs.~(\ref{n1})-(\ref{n3}) 
one concludes that the density of contact points corresponds to a local
operator with scaling dimensions $X = 1/3$. This result is surprising. One
can show that the profile of an operator with spin vanishes \cite{BUXU}]. This
implies $\Delta = \bar{\Delta}  = 1/6$. This value can't be obtained from the
possible values of $\Delta$ (see (\ref{A8})) of minimal models. It might be
obtained from $W(A_n)$ algebras \cite{flohr,bouwknegt}
 but it is not
clear why W-symmetry should play any role in our problem. Another
possible explanation is that for a conformal theory with $c = 0$
(this is also the case of percolation), under certain circunstances, 
 the scaling dimension can be an
arbitrary number \cite{bauer1,bauer2}. 
 This possibility is a bit strange since
the number one has to obtain is a neat $1/6$. Finally, it is possible, that
the proof that operators with spin have no profile functions does not apply
to our problem. If this is the case, one can choose $\bar{\Delta}  = 0$ and
$\Delta = 1/3$ 
\footnote{For the appearance of
operators with conformal spin in the presence of the quantum group symmetry
mentioned in Appendix A, see Ref.~\cite{ARHE}.}. 
 The value $1/3$ is a perfectly acceptable one (see Appendix A).
To conclude, we have not yet an explanation for the value $X = 1/3$ in
 Eq. (\ref{n3}).

%      We  show now how we can
%      derive this exponent using the results of Appendix A.
%
%       We have first studied the behavior of the density of contact points
%       $g(x,L)$ for
%       several lattice sizes and notice that for the function
%       where $y = x/L$ one has a data collapse. 
%
%%F7777777777777777777
%
%
%%F777777777777777777
%In the conformal invariance approach \cite{BUXU} we expect the scaling behavior
%\be \label{eq10}
%H(\cos(\pi x/L))  = [\frac{L}{\pi} \sin(\pi x/L)]^{\Delta + 
%\bar{\Delta}} g(x,L),
%\ee
%where $\Delta$ and $\bar{\Delta}$  are the scaling dimensions associated with a
%local operator of conformal spin $\Delta-\bar{\Delta}$.
%Using the scaling dimensions given
%in Appendix A we notice that one obtains the desired result taking $\Delta =
%1/3$ and $\bar{\Delta}= 0$ (see Fig.~\ref{fig6}(b)). Therefore the local density of contact points
%corresponds to a local operator of conformal spin 1/3 \footnote{For the appearance of
%operators with conformal spin in the presence of the quantum group symmetry
%mentioned in Appendix A, see Ref.~\cite{ARHE}.}. From  the value of $H(1)$ we 
%derive the one-contact point function in the thermodynamic limit 
%\be \label{eq11}
%g_{\infty}(x) = \frac{A}{x^{\frac{1}{3}}},
%\ee
% where the universal constant $A =0.758 \pm 0.005$ is close to the value 3/4.

\begin{figure}[t]
\centering{\includegraphics [angle=0,scale=0.59] {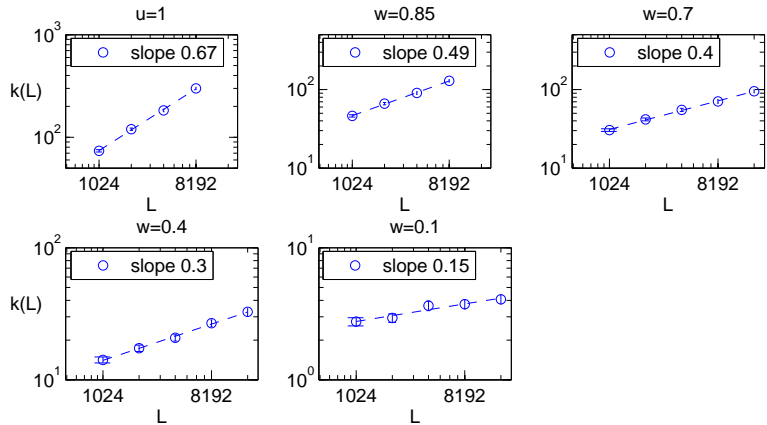}}
\caption{
   Average number of clusters $k(L)$ as a function $L$ for
  various values of~$w$ . The values of the exponent $\alpha$ 
(slope) is also shown.}
\label{fig8}
 \end{figure}

  We move away from $u =1$ and consider the domain $1 > w$. We observe that in  Fig.~\ref{fig8}
   the number of clusters $k(L)$ keep having an algebraic increase with an
  exponent $\alpha$ that  decreases monotonically with  $w$.
%F88888888888888888888888
% _________________________________________________________
%
%F888888888888888888888888
In Table~\ref{table1}   we give estimates for the exponent $\alpha$ for several values of
  $w$. We have also checked for several values of $w$ that the scaling law
\be \label{eq12}
G(x,L) = L^{1-\alpha} g(x,L) = G(\frac{x}{L})
\ee
  holds.

%T1111111111111111111
\begin{table}[t]
\centerline{
\begin{tabular}{|c|c|c|c|c|c|c|c|c|}
  \hline
  w & 1 & 0.85 & 0.75 & 0.4 & 0.25 & 0.1 & 0.05 & 0.025\\\hline
  $\alpha$ & 0.67 & 0.50 & 0.40 & 0.30 & 0.24 & 0.15 & 0.06 & 0.01 \\
  \hline
\end{tabular}
} \caption{
   Estimates of the exponent $\alpha$ giving the increase of the number
  of clusters with the size of the system $L$.}
\label{table1}
\end{table}
%T11111111111111111111

   From this investigation one can conclude that for values of $w > 0.05$ and
   maybe smaller, the system stays scale invariant with a varying critical
   exponent. In the next sections we are going to show that this   picture 
stays valid.

\begin{figure}[t]
\centering{\includegraphics [angle=0,scale=0.35] {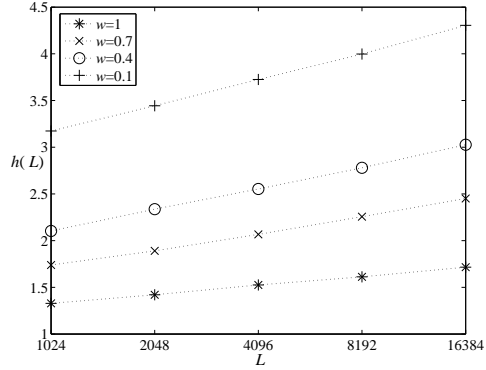}}
\caption{
      Average height $h(L)$ as a function of the
     lattice size
     $L$ for various values of $w \geq 0.1$.}
\label{fig9}
 \end{figure}

\begin{figure}[t]
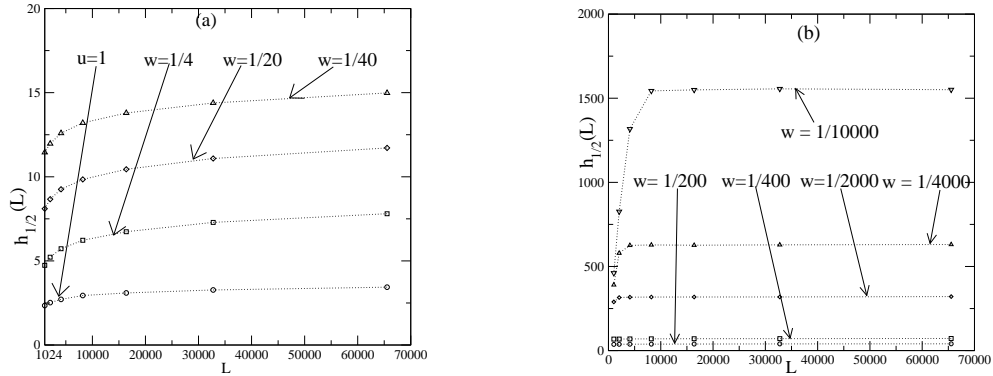

\centering{\includegraphics [angle=0,scale=0.35]
{sav19.eps}\hfill\includegraphics [angle=0,scale=0.35] {sav20.eps}}
\caption{
      (a) Average height in the middle of the lattice
     $h_{\frac{1}{2}}
     (L)$ as a function of the lattice size  $L$ for some  values
     of $w$ in the range
     $1 \geq w\geq 0.025$.
     (b)  Average height in the middle of the lattice
     $h_{\frac{1}{2}} (L)$
     as a function of the lattice size $L$ for various small values of $w$
     ($w \leq 1/200$).}
\label{fig10}\label{fig11}
 \end{figure}

    Although it seems obvious that one should look at the $L$ dependence of the
    average value of the heights $h(L)$ or the average value of the height at
    the middle of the lattice $h_{\frac{1}{2}}(L)$, it turns out that
    these quantities are
    harder to study since they vary slowly with the system size.
     In Fig.~\ref{fig9} we show $h(L)$ for various values of $w\geq0.1$.
     One sees a slow
     (logarithmic ?) increase when $L$ increases.
A similar picture is obtained if one considers $h_{\frac{1}{2}}(L)$
     (Fig.~\ref{fig10}(a)).

%F99999999999999999999999999

%     _______________________________________________________________
%     _________________________________________________________________

%F1010101010101010101010

     A  different picture emerges if one considers very small values of $w$ as
     shown in Fig.~\ref{fig11}(b). One can observe two different phenomena. The values of
     $h_{\frac{1}{2}} (L)$ increase substantially
     as compared to the values observed at larger values of $w$,
     and more interestingly,
     it looks like
     they saturate for large values of $L$.
     Two scenarios are compatible with our
     last observation. There is a change of the physics which occurs at very
     small values of $w$ illustrated by a change of the behavior of the $L$
     dependence of $h_{\frac{1}{2}}(L)$. This is the first scenario.
     If this would be the
     case, we would have observed a crossing of the increasing functions
     $h_{\frac{1}{2}}(L)$ obtained for larger values of $w$ and the flat
     values of $h_{\frac{1}{2}}(L)$
     observed for small values of $w$. This is not the case. What looks more
     plausible is  the second scenario where  for larger values of 
$w$, $h_{\frac{1}{2}}(L)$
     also saturates but this
     is seen only at larger values of $L$ than those considered in our 
simulations.

%F11-11-11-11-11-11-11

%F11-11-11-11-11-11-11

\begin{figure}[t]
\centering{\includegraphics [angle=0,scale=0.39] {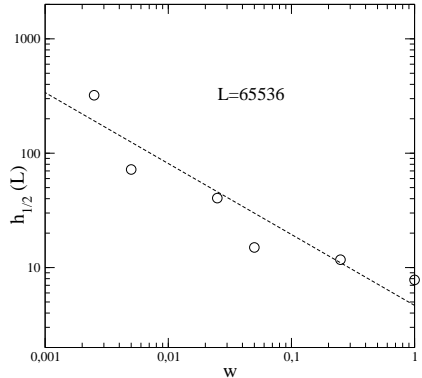}}
\caption{
      Average height in the middle of the lattice
     $h_{\frac{1}{2}} (L)$
     as a function of $w$ for the lattice size $L=65536$. The dashed curve is
     given by (\ref{eq13}).}
\label{fig12p}
 \end{figure}

\begin{figure}[th]
\centering{\includegraphics [angle=0,scale=0.49] {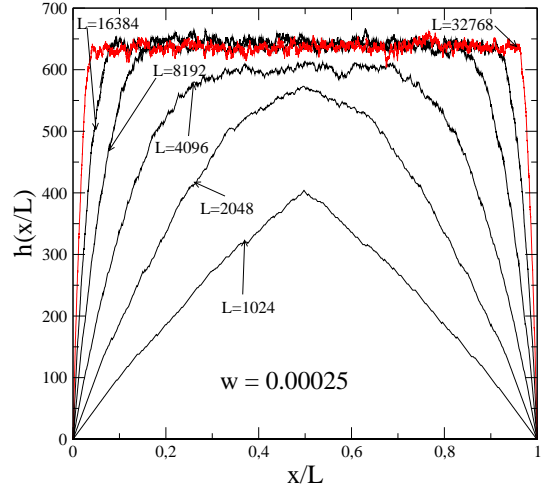}}
\caption{
Typical configurations in the stationary state for $w = 0.00025$ and
different values of $L$.}
\label{fig12pp}
 \end{figure}

According to the second scenario, for all values of $w<1$ (maybe
including $w = 1$) and very large values of $L$, $h_{\frac{1}{2}}(L)$ saturates. In order
to check if this scenario makes sense, taking $L=65536$ we have done a fit
to the curve which gives the maximum values of $h_{\frac{1}{2}}(L)$ (denoted by
$h_{\mbox{max}}$) for various values of $w$. One obtains:
\be \label{eq13}
 h_{\mbox{max}} = \frac{4.67}{ w^{0.62}}.
\ee
  As seen in Fig.~\ref{fig12p} one obtains a fair fit to the data.
  Interestingly,
  if we use (\ref{eq13}) for $w=1$ we obtain $h_{\mbox{max}} = 4.67$, a
 value compatible with
   the results shown in Fig.~\ref{fig10}(a).
  Taking the results shown in Fig.~\ref{fig11}(b) at face value, implies that the
  phase SOC (see Fig.~\ref{fig:phase1}) extends at least to the value
  $w = 10^{-4}$. On the
  other hand for $w=0$ the single configuration is the full triangle,
  $h_{\mbox{max}}$ has values of the size of the system and therefore the
  $L$-dependence of  $h_{\mbox{max}}$
  shows no saturation.
  This suggests a non-analytical behaviour of various quantities describing the system
  at $w = 0$. In order to illustrate this point, in Fig.~\ref{fig12pp}
  we show typical configurations
  in the stationary state for a very small value of $w$ ($w = 0.00025$) and
  different system
  sizes. One can see a
  clear change in the role of the boundaries if $L$ varies from 1024 to 32768.
  For $L = 1024$
  one is close to the full triangle configuration (expected for $w = 0$).
  However as one increases
  $L$ one gets a plateau (in which the height is almost constant and $L$
  independent) and the
  boundaries play a less and less important role.

%    _______________________________________________________________

%    _______________________________________________________________
%     ____

\section{ Space-time phenomena }
\label{sec4}
 In the last section, in which we have studied the properties of the
 stationary states, the data have confirmed the phase diagram shown in
 Fig.~\ref{fig:phase1}. In the present section we are presenting new facts which suggests
 that our assumption is correct.
  We are going to use the Family-Vicsek \cite{FAVI}
  scaling in order to check if
  the system is in a scale invariant phase and if the answer is
  positive, determine the dynamical critical exponent $z$.
   If we consider a time-dependent average quantity $a(t,L)$, being 
     $a(L)$ the average in the stationary state, than for large
   values of $t$ and $L$ the function $A(t,L)$ scales as:
\be \label{eq14}
A(t,L) = \frac{a(t,L)}{a(L)} -1 \sim A(\frac{t}{L^z}),
\ee
   where $z$ is the dynamical critical exponent. The scaling function depends on
   the initial conditions. In the data shown below, we have taken as initial
   condition the substrate ($h(2i) =0, h(2i+1)=1$, ($i=0,\ldots,n$))
   with probability one.
    We first consider the quantity $K(t,L)$ which corresponds to the
    average number of clusters ($a(t,L)$ in (\ref{eq14}) is $k(t,L)$ in this case).

\begin{figure}[t]
\centering{\includegraphics [angle=0,scale=0.33]
{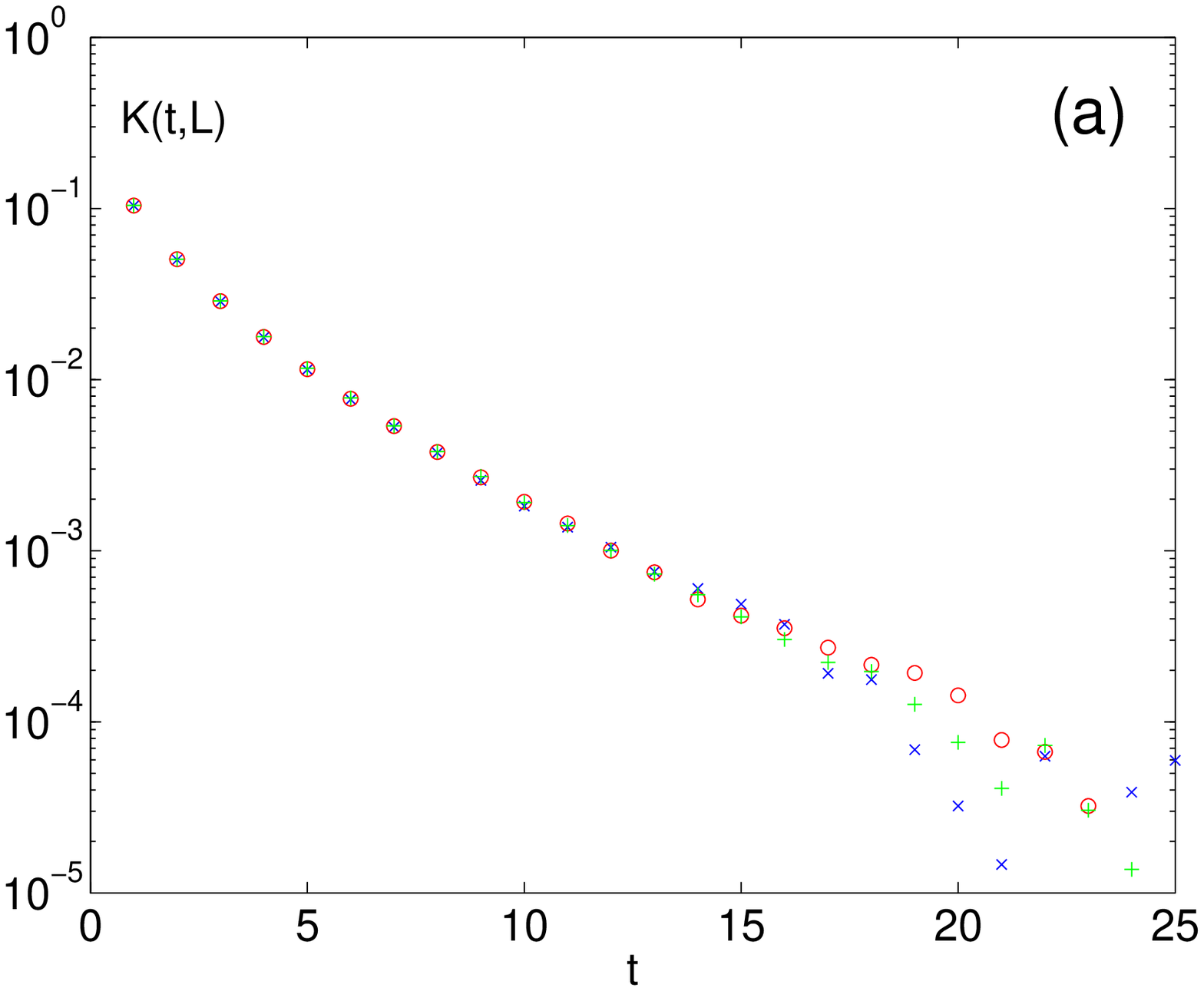}\hfill\includegraphics [angle=0,scale=0.44]{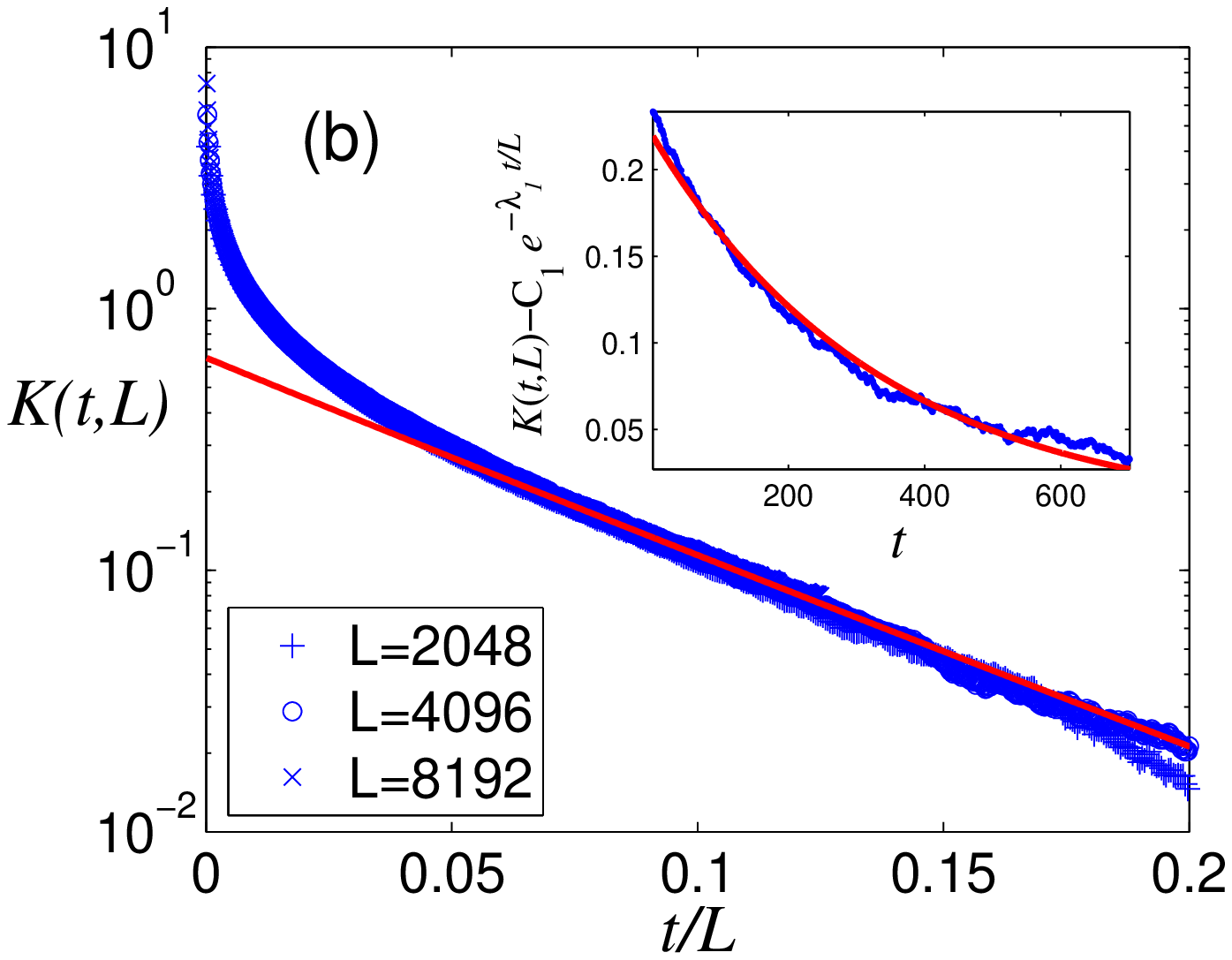}}
\caption{ (a)     Function $K(t,L)$, defined on (\ref{eq14}),
 for $u=0.4$ as a function of time for
    the lattice sizes $L=$ 2048, 4096 and 8192. 
    (b) Function $K(t,L)$ for $u=1$ as a
      function of $t/L$ for various lattice sizes. Red curve is given by a 
fit to (\ref{eq15}) considering only the first exponential  
with $\lambda_1=\pi \frac{9}{2}\sqrt{3}$ and $C_1 = 0.18$. In the inset 
 we compare $K(t,l)-C_1 e^{-\lambda_1 t/L}$ for $L=8192$ with the short time exponential $C_2 e^{-\lambda_2 t/L}$ (given in red line) with 
$\lambda_2= \frac{3}{2}\lambda_1$ and $C_2 = 0.54$.  }
\label{fig12}\label{fig13}\label{fig14}
 \end{figure}

 In   Fig.~\ref{fig12}(a) we show the function $K(t,L)$ for $u = 0.4$ for different lattice
    sizes.
%F12-12-12-12-12-12-12
%    ________________________________________________________________________
%    ______________________________________________________________________
%F12-12-12-12-12-12-12-12
We notice that for large lattice sizes, $K(t,L)$ is a function on $t$ only
     and can be obviously fitted with a sum of exponential functions as
     expected in a massive phase.
     Fig.~\ref{fig13}(b) shows $K(t,L)$ for $u = 1$ and as  expected from conformal
      invariance, $z=1$.
Conformal invariance gives also information on the function $K(t,L)$.
        One can use the knowledge of the finite-size scaling limit spectrum and obtain
       the functional dependence of $K(t/L)$.
      To illustrate how
      it works (see Fig.~\ref{fig14}(b)), for large values of t/L one can do a fit to the
      data, using the following ansatz:
\be \label{eq15}
K(t/L) = C_1 e^{-\frac{\lambda_1 t}{L}}
+  C_2 e^{-\frac{\lambda_2 t}{L}},
\ee
      where the $\lambda_i$ are obtained from the spectrum of the  Hamiltonian (\ref{A7}),
\be \label{eq16}
E_i = \frac{\lambda_i}{L}, \; i=1,2.
\ee
      The  constants $C_i$
      depend on the initial conditions. If one starts with the substrate 
      $C_1 = 0.28$ and  $C_2 =
      0.54$.

       We now consider smaller values of $w$ (larger values of $u$).
       We have done an
       extensive study of various functions $A(t,L)$ taking $k(t,L)$,
       $h(t,L)$ and
       $h_{\frac{1}{2}}(t,L)$ in  (\ref{eq14}).
       The estimated values for the critical exponent $z$
       are shown in Table~2. For very small values of $w$, the estimates
       coming from $K(t,L)$ are poor because the number of
       clusters is small and hence
        large errors.
%
%
%T2222222222222222222222
\begin{table}[t]
\centerline{
\begin{tabular}{|c|c|c|c|}
\hline
$w$ & $K$ & $H$ & $H_{1/2}$\\
\hline
1 & 1.03 & 0.95 & 0.97  \\
0.9 & 0.86 & 0.74 & 0.72 \\
0.7 & 0.55 & 0.52 & 0.50 \\
0.4 & 0.35 & 0.34 & 0.34 \\
0.25 & 0.31 & 0.30 & 0.30 \\
0.025 & * & 0.07 & 0.07 \\
\hline
\end{tabular}}
\caption{
       Estimates of the dynamical critical exponent $z$ 
for various values of $w$. The
       estimates were obtained using  (\ref{eq14}) with $A$ replaced by  $K$, $H$ and $H_{\frac{1}{2}}$.   The number represented by (*) is not reliable since in that region the
       number of clusters is very small.}
\label{table2}
\end{table}
%
%
%      _________________________________________________________________
%       table 2
%     ____________________________________________________________________

\begin{figure}[t]
\centering{\includegraphics [angle=0,scale=0.35] {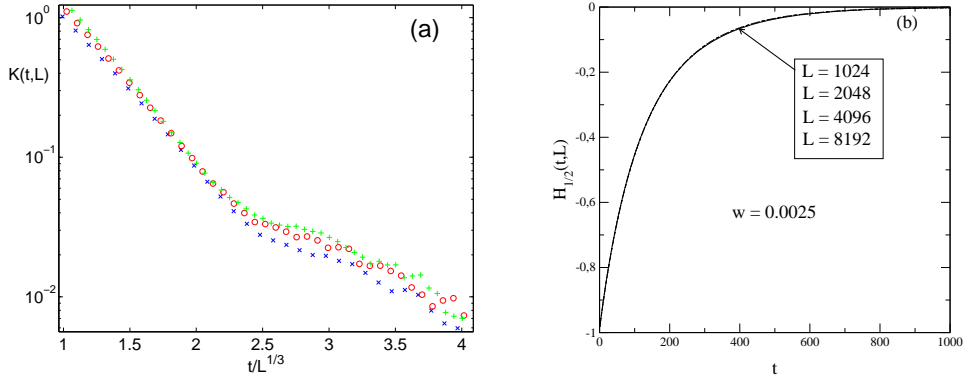}\hspace{1.0cm}\includegraphics [angle=0,scale=0.37] {sav15.eps}}
\caption{
(a) $K(t,L)$ for $w=0.4$ as a function of
    $t/L^{\frac{1}{3}}$ for the lattice sizes $L = 1024,2048,4096$ and 8196.
    (b) $H_{\frac{1}{2}}(t,L)= h_{\frac{1}{2}}(t,L)/h_{\frac{1}{2}}(L) -1$ for $w=0.0025$
    as a    function of $t$ for various lattice sizes.
    }
\label{fig15}\label{fig16}
 \end{figure}

%T2222222222222222222222
%
        Inspection of Table~\ref{table2} shows a smooth drop of the values of
    $z$  from $z=1$ for $w=1$ to $z=0$ for $w=0$. The
    latter values is in agreement with a direct calculation of the
    mass gap at $w=0$ \cite{GNPR} which gives a finite gap in the
    thermodynamic limit.
    To show how the estimates given in Table~\ref{table2} work, we demonstrate in Figs.~\ref{fig15}(a) and \ref{fig15}(b)  the scaling
    functions $K(t,L)$ for $w = 0.4$ and $H_{\frac{1}{2}}(t,L)$
    for $w  = 0.0025$, respectively.

%F15-15-15-15-15-15-15

%______________________________________________________

\begin{figure}[t]
\centering{\includegraphics [angle=0,scale=0.39] {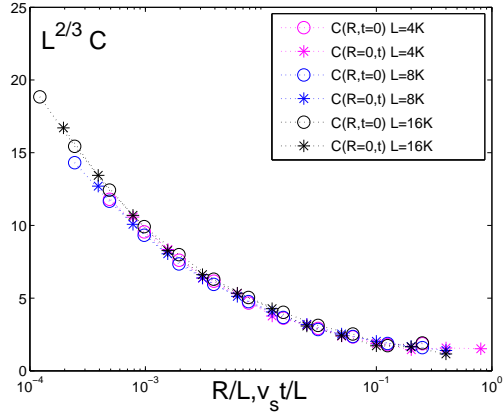}}
\caption{
       The 2-contact point correlation function
      $C(R,t,L)$
      times $L^{\frac{2}{3}}$ as a function of the scaled distance
      $R/L$ and the scaled time
      $\frac{v_s}{L}t$ ($v_s = \frac{3}{2}\sqrt{3}$ is the sound velocity)
      for various lattice sizes.}
\label{fig17}
 \end{figure}

     We have not studied extensively the two-contact points
 space-time correlation functions
      except
      for the special case $u = 1$. This case is interesting since 
from conformal invariance 
       the two-contact points space-time  correlation function exhibits, 
far from the  boundaries, 
       the following scaling form in the bulk (see (\ref{eq9})):   
\be \label{eq18}
C(R,t,L) = L^{-\frac{2}{3}} G(\mu),
\ee
      where
\be \label{eq19}
\mu = \frac{\sqrt{R^2 +v_s^2t^2}}{L}.
\ee
      In (\ref{eq18}) and (\ref{eq19})  $R$ is the distance between the contact
      points, $t$ is the time difference and $v_s= \frac{3}{2}\sqrt{3}$ is
      the sound velocity (see Appendix~A).
      The factor $L^{-\frac{2}{3}}$ in (\ref{eq18}) is obtained from  (\ref{eq9}).
      In Fig.~\ref{fig17} we show the scaling function $G(\mu)$
      \footnote{We have not used the
      connected correlation function in (\ref{eq18}).}.
      To avoid boundary  effects the data was taken only in the center segment of size $L/2$.

%F17-17-17-17-17-17

%     _______________________________________________________________
%    _________________________________________________________________

%F17-17-17-17-17-17

       The functional dependence (\ref{eq18}) which is special for the conformal
       invariant case is nicely seen in Fig.~\ref{fig17}.
       Our data are not good enough in
       order to extract from the small $\mu$ behavior of $G(\mu)$,
       the two-contact
       point correlation function in the thermodynamic limit.
%*********************************************

\section{  Avalanches in the RPM}
\label{sec5}
 An interesting property of the RPM is the occurrence of avalanches in the
 desorption processes. For the case $u = 1$, based on exact results on
 small lattices, the existence of avalanches was already suggested in
 \cite{GNPR}.
 In this section, using  Monte Carlo simulations on large
 lattices, we  present the properties of avalanches in the whole
  domain $0 < w \leq 1$.

  In the stationary state, once a tile from the rarefied gas hits the
  interface, it can be reflected, adsorbed or can trigger a nonlocal
  desorption process in which many tiles may leave the interface, this defines 
an avalanche. In the latter
  case the size of the avalanche is given by the number of tiles $T$ that 
are released in the process. 
  This number is
  always odd, therefore it is convenient to write: $T = 2v -1$
  ($v =1,2,\ldots$). For $u<1$ only a finite number of tiles are
  removed, since the density of clusters is finite. For $w\leq 1$, the
  cluster density vanishes, and therefore macroscopic number of
  tiles may be desorbed, hence  macroscopic avalanches.

   We denote by $S(v,L)$ the PDF which gives the probability for an
   avalanche
   of size $v$ for a system of size $L$. In models of self-organized
   criticality
   (SOC) \cite{BTW,RRA,DHA,TMS,PIP}, for large values of $v$ and $L$,
   one may expect
   the PDF to exhibit a simple scaling law,
 \be \label{B.1}
 S(v,L) = v^{-\tau} F(\frac{v}{L^D}),
 \ee
   characterized by two exponents $\tau$ and $D$.

   It is convenient to consider moments of the PDF and their scaling
    properties:
\be \label{B.2}
<v^m>_L = \sum_{v=1}v^mS(v,L) = \Gamma_mL^{\sigma(m)}
\ee
    where
    \be \label{B.3}
    \sigma(m)=
    \left\{ \begin{array}{cc}
    0, & m<\tau-1 \\
    D(m+1-\tau), & m > \tau-1
    \end{array}\right. .
    \ee

     In order to get estimates for $D$ and $\tau$ from the Monte Carlo
     simulations,
     it is convenient to take couples of system sizes $L$ and $L'$ and consider
     the quantities:
\bea \label{B.4}
\sigma_{L,L'}(m) = \ln \large(\frac{\langle v^m \rangle_L}{\langle v^m
\rangle_{L'}}\large) / \ln\large(\frac{L}{L'}\large) = m D_{L,L'}  + B_{L,L'}
 \eea
     and
\be \label{B.5}
\sigma_{L,L'}(m_{L,L'}) = 0, \; \; \; \; \tau_{L,L'} = m_{L,L'} + 1.
\ee

\begin{figure}[t]
\centering{\includegraphics [angle=0,scale=0.49] {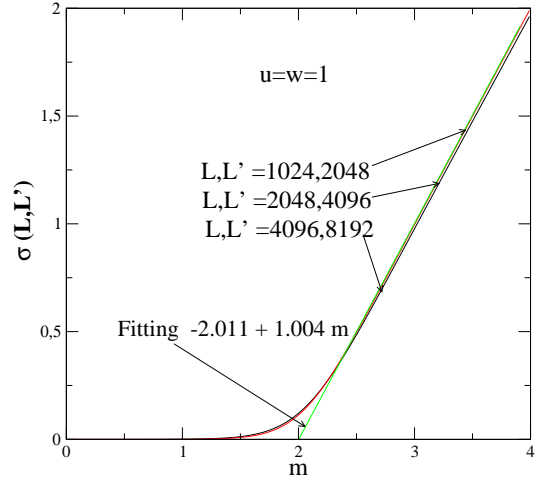}}
\caption{
     The $m$ dependence of $\sigma_{L,2L}$ for $w=1$. The fitted straight line
is also shown. }
\label{fig19}
 \end{figure}

      Using the results of Monte Carlo simulations for couple of sizes (in
      practice we have chosen $L' = 2L$) one can obtain estimates for the two
      critical exponents $D$ and $\tau$ from 
$\sigma_{L,2L}$ and $\tau_{L,2L}$.
       In Fig.~\ref{fig19}  we illustrate the $m$
       dependence of $\sigma_{L,2L}$
       obtained from Monte Carlo simulations, for several values of $L$ in the
        $u = 1$  case.
%_______________________________________________________________________________
%
%
%____________________________________________________________________

\begin{figure}[t]
\centering{\includegraphics [angle=0,scale=0.45] {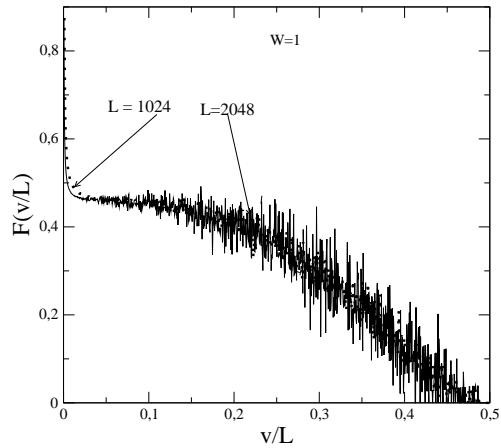}}
\caption{
The scaling function $F(v/L)$ for $w=1$. The data are for lattice
sizes $L=1024$ and $L=2048$.}
\label{fig10-1}
 \end{figure}

%     _______________________________________________________________
%    _________________________________________________________________
 From a cursory look at this figure  one can see that for
$u = 1$, $\tau \approx  3$ and
    $D \approx 1$ confirming the results obtained in \cite{GNPR}
    using much smaller lattices
    (for a more precise estimate, see Table~\ref{table3}). The result $D=1$ is to be
    expected since in a conformal invariant theory (this is the case if
    $u = 1$) one has no other length than the size of the system $L$.
    We found no
    explanation for the value three of $\tau$ although this number
    shows up in
    many aspects of the model (see for example  (\ref{eq9}) and the
    Appendix).
    In Fig.~\ref{fig10-1} we show the scaling function $F(v/L)$
    defined in  (\ref{B.1}). One observe a nice data collapse.

     In Table~\ref{table3} we give the estimates for the exponent $\tau$ and $D$
     for various
     values of $w$. The estimates presented in this Table are obtained using only
     one pair of sizes (4096 and 8192) but we made  sure that the
     results are
     reliable  by studying the changes of the values of the estimates
     using smaller lattices.
%_______________________________________________________________________________
%
%
%____________________________________________________________________

\begin{figure}[t]
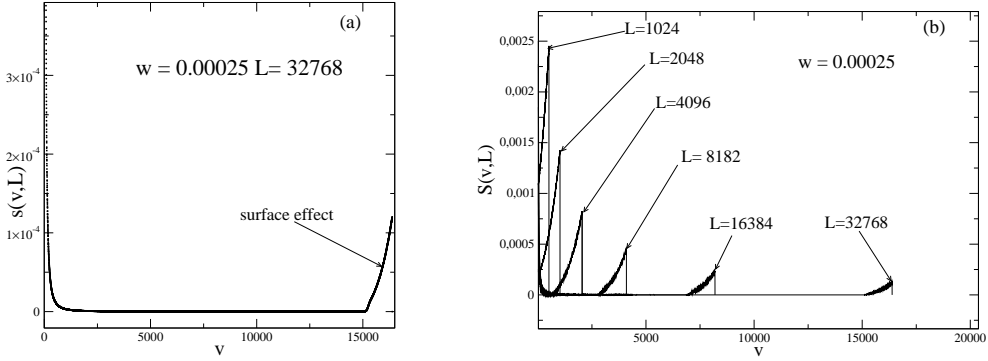

\centering{\includegraphics [angle=0,scale=0.37] {tave5.eps}\hfill
\includegraphics[angle=0,scale=0.37] {tave6-1.eps}}
\caption{
(a) The PDF for avalanches for $w = 0.00025$ and $L = 32768$.
(b)  The PDF of avalanches for $w = 0.00025$ and different sizes.}
\label{fig10-2}\label{fig10-3}
 \end{figure}

\begin{figure}[t]
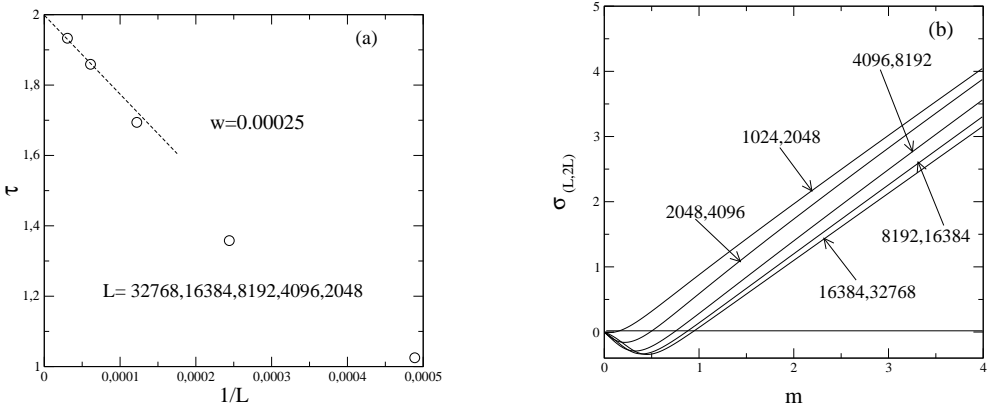

\centering{\includegraphics [angle=0,scale=0.40]
{tave10.eps}\hfill\includegraphics [angle=0,scale=0.40] {sav40.eps}}
\caption{ (a)  Estimates $\tau_{L,2L}$ for $w = 0.00025$ using
different values of
  $L$ from
  $L =  2048$ to $L = 16384$
(b)  The $m$ dependence of $\sigma_{L,2L}$ for $w=0.00025$. }
\label{fig10-4}\label{fig20}
 \end{figure}

For small values of $w$, the determination of $\tau$ has to be done with care.
This is due
to surface effects (there are avalanches occurring at both ends of the system which have
a different behaviour than avalanches occurring in the bulk). The phenomenon is
illustrated in Fig.~\ref{fig10-2}(a) where the PDF $S(v,L)$ is shown for
$w = 0.00025$ and $L = 32768$.
%_______________________________________________________________________________
%
%
%____________________________________________________________________
One notices that $S(v/L)$ has two peaks: one for $v = 1$ and another one
for very large
values of $v$.  Large avalanches  occur when one tile hits the boundary
of the interface.
On the other hand, one expects that for large values of $L$,
the probability for a tile
to hit a boundary  get smaller and smaller as the system gets larger
(see Fig.~\ref{fig12pp}),
therefore the probability to have large avalanches should decrease with the size of the
system. This phenomenon is nicely illustrated in Fig.~\ref{fig10-3}(b),
where one can see that the probability
to have an avalanche at the boundary decreases with the size of the system.
 One expects that the determination of $\tau$ using
  (\ref{B.4})    and (\ref{B.5}) will give a smaller
 value for $\tau$ for small lattices (large avalanches at the boundaries)
 and a larger
 value when compared with the value obtained by using large  lattices 
  (the importance of the boundary effects decreases).
 That
 is exactly what happens as can be seen in Fig.~\ref{fig10-4}(a).
 One sees in this figure  that the estimates for $\tau$
 converge to a value close to  2.
%_______________________ma________________________________________________________
%

      In order to illustrate the convergence of the estimates in
      Fig.~\ref{fig20}(b) we
      present  for a very small value of $w$ the
      functions $\sigma_{L,2L}$ for several values of
      $L$.
Although the convergence is poorer when compared with these quantities 
obtained for 
higher values of $w$ (see Fig.~\ref{fig19}), one can still get reliable results.
%T3333333333333333333333333
\begin{table}[t]
\centerline{
\begin{tabular}{|c|c|c|c|c|c|c|c|c|c|c|}
  \hline
  $w$ & 1 & 0.95 & 0.90 & 0.80 & 0.60 & 0.45 & 0.30 & 0.05 & 0.02 & 0.005\\
  $D_{L,L'}$ & 1.004 & 1.089 & 1.087 & 1.066 & 1.040 & 1.026 & 1.017 & 1.002 & 1.002 & 1.006 \\
  $\tau_{L,L'}$ & 3.00 & 2.77 & 2.63 & 2.47 & 2.32 & 2.25 & 2.18 & 2.07 & 2.046 & 2.00 \\
  \hline
\end{tabular}}\caption{
      Estimates of the critical exponents $D$ and $\tau$ obtained
      by using lattices sizes
      $L = 4096$ and $L' = 8196$.}
\label{table3}
\end{table}

%        Remarkably the value of the exponent $D$ stays unchanged and
%        equals one in
%        the whole range $w<1$. In a quite different context, a neat
%     mean-field calculation in a very simple model \cite{STKC}
%     gives a similar
%     result: $\tau$ changes while $D$ stays fixed.
%     In our case, the result is a bit
%     mysterious: for $w=1$ the value $D=1$ is easily
%     understood, but the system has a very different space-time symmetry
%     for
%     other values of $w$. This is not the case for the model studied in
%     Ref.~\cite{STKC} where $D = 2$ and the process are governed by the
%     diffusion equations.
%  The robustness of the exponent $D$ is probably of a more
%general validity than our model.

The value of the exponent $D$ stays unchanged and equals to one in the whole 
range $w < 1$. This can be understood in the following way 
\cite{dhar-private}
 the tiles which are desorbed and create the avalanche 
belong always to a one-dimensional layer.

\begin{figure}[t]
\centering{\includegraphics [angle=0,scale=0.39] {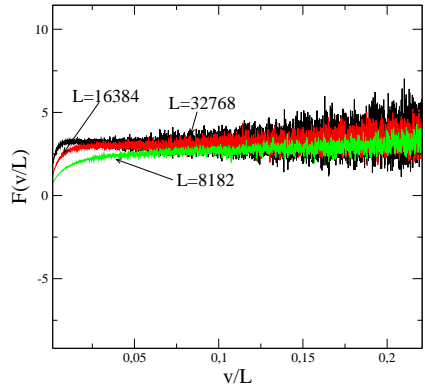}}
\caption{
 The functions $F(v/L)$, defined on (\ref{B.1}),
 at $w=0.00025$  for some values of $L$. }
\label{fig20p}
 \end{figure}

 We turn now to the exponent $\tau$. As can be seen in Table~\ref{table3} this exponent
varies from the value 3 ($u = 1$) to a value close to 2 for $w$
close to zero. In the case $u = 1$, the average size of the
avalanches stays finite when $L$ gets large but the dispersion
diverges (the function $F(v/L)$ shown in Fig.~\ref{fig10-1} does not
vanish at the origin). Does an exponent $\tau = 2$ mean that the
average size of the avalanches  diverges logarithmically with $L$ ?
Not necessarily, it depends if the scaling function $F(v/L)$
vanishes or not at the origin. In Fig.~\ref{fig20p} we show for $w =
0.00025$, that  $F(v/L)$ does not vanish at the origin. Since later in
this section we are going to show that $\langle v \rangle$ stays
finite for any $w >0$, we conclude that $\tau$ gets very close to
the value 2 but never reaches it.

 We would like to comment now on the properties of the avalanches seen at the
boundaries and illustrated in Fig.~17.
 For large values of $v$ where the
boundary effects occur, the PDF $S(v,L)$ presents two striking
features: the probability to have an large avalanche decreases with
$L$ and $S(v,L)$ increases with $v$. A close inspection of the local
heights distribution for different values of $L$ seen in
Fig.~\ref{fig12pp} allows us to investigate both features. Firstly,
one sees that for large values of $L$ the boundary becomes less and
less important and therefore the probability for a tile to hit the
boundary and produce a large avalanche decreases. On the other hand,
at the boundary, the probability for a hit by a tile is proportional
to $|dh(x,L)/dx|$ (keep in mind that a desorption process takes place
when the local slope $s(i)$ takes the values $\pm 1$). From the data
shown in Fig.~\ref{fig12pp} one can see that for $x$ close to the
boundaries, $d^2 h(x,L)/dx^2$ is negative. This implies that the
probability to hit the interface with a tile at the two ends of 
the system  is
larger than further into the system.
%-------------------------------------------
%

An important consequence of this study is that avalanches exist at least 
in the interval $1 \geq w \geq 0.00025$ and therefore the system is scale 
invariant in this interval.(The data shown in Tables \ref{table1} and 
\ref{table2} are limited 
to the interval $1 \geq w \geq 0.025$). It is probably safe to assume 
that the 
SOC phase covers the domain $1 \geq w \geq 0$. (We have 
 obviously checked that for $u<1$,
      where the
      system is massive, for large values of $v$ the PDF has an exponential
      decrease and not an algebraic one).

       We turn now our attention to the average size of the avalanches. A
       simple mean-field argument allows to compute it using informations
       about
       the stationary state only.

        We first consider, in the stationary state, the average fraction of
         interface where desorption does not take place $n(L)$ (FND)
         defined in
        (\ref{m}). This quantity   varies from the
        value 1 for the substrate
        (the single
        configuration for $u = 0$) to the value $1/(L-1)$ for the full
        triangle (the
        single configuration for $w = 0$).

         For $u = 1$ there is a conjecture \cite{STRO} for the values
         of $n(L)$
\be \label{B.6}
n(L) = \frac{3L^2-2L+2}{(L-1)(4L+2)} =\frac{3}{4}(1-\frac{1}{6L}+\cdots ).
\ee
          This conjecture was checked using our Monte Carlo simulations
          for
          various lattice sizes.

           If one knows $n(L)$, the average probability $P_a(L)$ to have
           an adsorption
           process, the average probability $P_r(L)$ to have a reflection
           process and
           the average probability $P_d(L)$ to have a desorption process
           can be easily
           obtained:
\be \label{B.7}
P_a(L) =  \frac{n(L)}{  2} - \frac{1}{2(L-1)},
\ee
\be \label{B.8}
P_r(L) = \frac{n(L)}{2}(2w-1) +1-w + \frac{1}{2(L-1)},
\ee
\be \label{B.9}
P_a(L) + P_r(L) + P_d(L) = 1,
\ee
where, like in our Monte Carlo simulations, we choose $u_a =1$ and
$u_d=w=1/u$. 

            Taking into account that in the stationary state the average
        number of
        adsorbed tiles is equal to the average number of desorbed
        tiles, one
        obtains:
\be \label{B.10}
\langle T\rangle_L = 2\langle v\rangle_L -1 = \frac{P_a(L)}{P_d(L)}.
\ee
Notice that this is a mean-field calculation since we have first computed 
the average probability to have a desorption process and multiplied it 
with the average number of tiles which are desorbed.
       In the large $L$ limit (\ref{B.7})-(\ref{B.10}) gives:
\be \label{B.11}
\langle T \rangle_{\infty} \approx \frac{n_{\infty}}{2(1-n_{\infty})w}.
\ee

\begin{figure}[t]
\centering{\includegraphics [angle=0,scale=0.39] {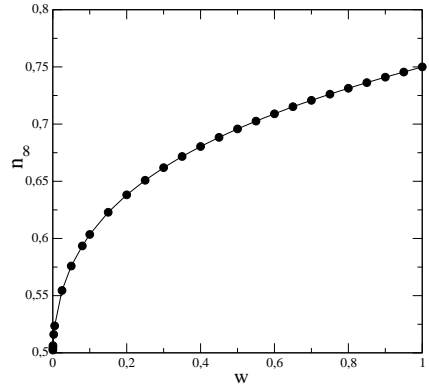}}
\caption{
   The $w$ dependence of the fraction of the interface where
 adsorption does not take place $n_{\infty}$ obtained from
the extrapolated results of Monte Carlo
 simulations. }
\label{fig21}
 \end{figure}

          In Fig.~\ref{fig21} one shows the $w$ dependence of
          $n_{\infty}$ as obtained from
          the extrapolated results of our Monte Carlo simulations.
One notices a discontinuous behavior of $n_\infty$ around $w=0$. At
$w=0$ one has $n_\infty = 0$, while $ \lim_{w \to 0_+} n_{\infty} =  0.5$.
  Using (\ref{B.11}) one concludes,
in agreement with previous observations that for any finite value  $w$, the
average size of the avalanches is finite.

 In Table~\ref{table4}  we compare in two cases the mean-field predictions for
$\langle T\rangle_L$ given in
(\ref{B.11}) with the values measured in Monte-Carlo simulations. The agreement is
excellent. In the same Table the values for $n(L)$ for several values of $L$ are also
given. One notices that for $u = 1$ and $L = 1024$ one  already  gets within five digits the
asymptotic value 3/4. This is not the case for $w = 0.00025$, where 
the finite-size effects are larger. 
Our numerical analysis indicate that the asymptotic behavior 
of $n(L)$ can be described by
\be \label{B.12}
 n(L) = n_{\infty} ( 1 - A(w)/L + \cdots),
\ee
where $A(1) = 1/6$ and $A(w)$ diverges if $w \rightarrow 0$ (data not shown).

 We conclude this section with the following observation, which is necessary for the consistency of
the present
model for avalanches. If one looks at the expression (\ref{B.8}) of the average
(over configurations) probability that a tile hitting the interface is reflected,
one sees that for small values of $w$, the probability approaches the value one. This
implies that to obtain large avalanches (small $\tau$) one has to wait a long
time. This is to be expected in models of SOC.

%
%T444444444444444444444444444444444
\begin{table}[t]
\centerline{
\begin{tabular}{|c|c|c|c|c|c|}
\hline
 $w$ &$L$ & $n_L$ & $\langle T\rangle_L$ &MF&
 $ \langle T\rangle_L/\mbox{MF}$   \\
\hline
1 & 1024 & 0.749  & 1.497 & 1.497 & 0.999 \\
1 & 2048 & 0.749  & 1.498 & 1.501 & 0.998 \\
1 & 4096 & 0.749  & 1.499 & 1.500 & 0.998\\
1 & 8192 & 0.749  & 1.499 & 1.499 & 0.999 \\
\hline
\hline
0.00025 & 1024 & 0.199  & 496.999 & 496.119 & 1.001 \\
0.00025 & 2048 & 0.313  & 914.346 & 913.820 & 1.001\\
0.00025 & 4096 & 0.407  & 1374.713 & 1373.856 & 1.001 \\
0.00025 & 8192 & 0.455  & 1674.877 & 1674.550 & 1.000 \\
\hline
\end{tabular}}
\caption{
            Values of $n_{\infty}$ and  the average number of
            tiles obtained from Monte
            Carlo simulations $\langle T\rangle_L$ as compared
            with the compared with the mean-field
            results (MF). The values are given for several lattice
            sizes $L$ and two
            values of~$w$, $w=1$ and $w = 0.00025$.}
\label{table4}
\end{table}
%T44444444444444444444444444444444
%

\section{  Discussions}
\label{sec6}

The raise and peel model describes a stochastic process of a
one-dimensional interface. RSOS paths give the configuration space.
It is well known that if a stochastic model
has as stationary state in which the RSOS paths have the same
probability, the average height $\langle h(L)\rangle$ increases with the system
size $L$ like $\sqrt{L}$ (Brownian motion). In the one-parameter
($u$ or $w = 1/u$) RPM, in the stationary states, the various RSOS paths get
weights and henceforth the physics is different. It turns out that for
all finite values of $u$ ($w \neq 0$) in the large $L$ limit, $\langle h(L)\rangle$ stays finite.
 The convergence to the large $L$ limit of $\langle h(L) \rangle$ is very slow. The $w
= 0$ value is singular: in the stationary state one has, with
probability one the full triangle and therefore $h = L/2$.
 The phase diagram is shown in Fig.~\ref{fig:phase1}. For $u < 1$, one is in the
massive phase (finite correlation lengths and non-vanishing densities
of clusters). For $1 \geq w > 0$ the density of clusters vanishes
algebraically with an exponent which depends on $w$  (Table~\ref{table1}). The
dynamical exponent $z$ also depends on $u$. It  varies between one and zero
 (see Table~\ref{table2}). We called this phase the SOC phase since in the desorption
process one has avalanches. The PDF of the number of tiles desorbed
has long tails. Although the average number of tiles involved in the
avalanches stays finite for large $L$, higher moments diverge. The
$\tau$ exponent changes from the value 3 ($w = 1$) to the value 2
for $w$ close to zero (Table~\ref{table3}). The exponent $D$ is
robust and keeps the value $D = 1$. 
 A fascinating aspect of the model is that, contrary to our 
intuition, for any small value $w > 0$, if $L$ is
large enough, the system separates into a bulk part and a "surface"
part related to the condition that at the two boundaries the
heights have to take the value zero
(see Fig.~\ref{fig12pp}).

 The value $u = 1$ is special. In the continuum time limit the
Hamiltonian which gives the Markovian time evolution of the
stochastic process, has the same spectrum as the XXZ quantum spin 1/2
chain Hamiltonian with a given $Z$-anisotropy and fixed boundary terms
(see Appendix A). This Hamiltonian is integrable and the finite-size
scaling spectrum is known. The main observation is that for $u = 1$ one has
a space-time symmetry (conformal invariance). There are several
important consequences concerning the properties of the system. The
one-point function which gives the local density of contact points
has the scaling form given by  (\ref{n3}).
 The two-contact points
space-time correlation function has a scaling form given by  
 (\ref{eq18}). The dynamical
critical exponent $z$ can be computed analytically and has the value 1.
 There are other consequences of conformal invariance. The
Family-Viczek scaling function which gives the time and size
dependence of average quantities can also be obtained  (see Fig.~\ref{fig14}(b)).
The value $D = 1$ is a consequence of the fact that in a conformal
theory, in the large $L$ limit the single scale is $L$ itself. The value
$\tau = 3$ remains a puzzle although the number 3 appears all over for $u = 1$
(see Appendix A).

 The reader not familiar with conformal invariance might skip the
end of this section.
 We have shown that the density of contact points
gives a local operator of a $c=0$ ($c$ is the central charge of the Virasoro
algebra) conformal field theory. 
As discussed in Section 3 the value of the scaling dimension of
this operator has not a simple explanation. 
 Other local operators can be
defined in a similar way. For example one can look at the local density
of points where one has a deep (slope zero) at the height 1 (or any fixed
height).

 One interesting aspect of the RPM is that one has a simple
 example of an unusual kind of conformal symmetry breaking. 
For $u < 1$ the system
becomes massive. This is not a surprise. One can understand why this
happens in the following way: one can take the desorption rate $u_d =
1$ (see Section~\ref{se:modeldef}) and $u_a = u$. This implies that the perturbation
is local (the adsorption is local). If $w < 1$ the situation
is quite different and here comes the surprise. One has scale
invariance without having conformal invariance. The appearance of
this novel phenomenon has the following explanation. For $w < 1$  one takes
$u_a = 1$ and $u_d = w$. Desorption is a nonlocal process and therefore
the conformal theory is perturbed nonlocally and there is no a priori
reason why conformal invariance cannot be broken while scale
invariance is not.

\section*{Acknowledgements}
We would like to thank the referee for pointing out an embarrassing mistake
in Section 3. 
It is our pleasure to acknowledge   J. L. Cardy, B. 
Schittmann and  D. Dhar for reading this manuscript and useful discussions. 
We also thank M. Bauer, J. de Gier and M. Flohr for related discussions.   
FCA research was supported in part by FAPESP and CNPq (Brazilian Agencies).
Research of EL was supported by the NSF PFC-sponsored Center for Theoretical Biological Physics (Grants No. PHY-0216576 and PHY-0225630).
VR gratefully acknowledges support from the Deutsche Forshungsgemeinschaft,
The Australian Research Council, the EU network HPRN-CT-2202-00325 and FAPESP
(Brazil).

       \appendix

\section{ The finite-size scaling limit of the spectrum of the Hamiltonian
for $u=1$. Results from conformal field theory.}

 As shown in Ref.~\cite{GNPR}, in continuum time, the time evolution of the RPM
 for $u = 1$, is given by the Hamiltonian:

\be \label{A1}
H = \sum_{i=1}^{L-1} (1- e_i),
 \ee
 where $e_i$ acts on the configurations (RSOS paths) described in Section~1, taking one path into another or leaving it unchanged.
 Using a similarity transformation from the configuration basis to a spin basis, the $e_i$'s have a
 representation in terms of Pauli matrices:
\be \label{A2}
e_i = \frac{1}{2}[
\sigma_i^x\sigma_{i+1}^x +
\sigma_i^y\sigma_{i+1}^y -\frac{1}{2}
\sigma_i^z\sigma_{i+1}^z + \frac{1}{2}
+i\frac{\sqrt{3}}{2} (\sigma_i^z - \sigma_{i+1}^z)].
\ee

  If we introduce the expressions (\ref{A2}) in  (\ref{A1}),
  one obtains a well known
  Hamiltonian of an integrable quantum chain acting into a $2^L$  dimensional
  vector space. This chain has $U_q(sl(2))$ symmetry with $q = \exp(i\pi/3)$.
   To obtain the action of $H$ in the vector space of the RSOS paths one has
   to consider only the subspace of $U_q(sl(2))$ scalars (spin $s=0$)
 which has the
   dimension $C_n = (2n)!/(n+1)(n!)^2$  with $L=2n$.
    The spectrum of the Hamiltonian (\ref{A1}) is non-negative. Let us denote by
    $E_r$ ($r=0,1,\ldots $) the energy levels in increasing order:
\be \label{A3}
E_0 =0 <E_1<E_2<\cdots .
\ee
     The partition function giving the finite-size limit of the spectrum of
     the Hamiltonian of the quantum chain, is defined as follows:
\be \label{A4}
Z(q) = \lim_{L \rightarrow \infty} Z_L(q) =
\lim_{L\rightarrow \infty} \sum_{n=0} q^{\frac{LE_n}{\pi v_s}},
\ee
      where $v_s= 3\sqrt{3}/2$.
        One can show \cite{BASA} that $Z(q)$ has the expression
\be \label{A5}
Z(q) = \sum_{s=0}^{\infty} (2s+1)\xi_s(q),
\ee
    where
\be \label{A6}
\xi_s(q) = q^{\frac{s(2s-1)}{3}} (1-q^{2s+1}) \prod_{n=1}^{\infty}
(1 - q^n)^{-1}
\ee
    is the partition function in the spin $s$ sector.

      In Section~4 we look at the time evolution of the
  stochastic model. The finite size-scaling limit is given by (\ref{A6})
  with
  $s=0$. This implies that in the large $L$ limit we have
\be \label{A7}
LE_0 = 0, \; \; \; \; \;
LE_1 = \pi 3\sqrt{3},\;\;\;\;\;\;
LE_2 = \pi \frac{9}{2}\sqrt{3}, \;\;\;\;\;\ldots .\;\;.
\ee

 Up to now we have described the time evolution of the system. If one
considers space-time correlation functions, the various critical exponents
are expected to be obtained \cite{ab2,alcagrim} from the scaling dimensions of the primary
fields of minimal models \cite{itzykson}
 with a central charge c=0 [1]:
\be \label{A8}
\Delta = \frac{(3p-2q)^2 - 1}{24},
\ee      
where $p$ and $q$ are positive integers.
Notice that the value $\Delta = 1/3$ appears in both the finite-size spectrum
of $H$ (see (\ref{A6}) with $s = 1$) and in (\ref{A8}).

\end{document}